\documentclass[twocolumn,showpacs,amsmath,aps]{revtex4}
\usepackage{graphicx,color}
\usepackage{CJK}

\newcommand{\nl}{\nonumber \\}

\newcommand{\be}{\begin{equation}}
\newcommand{\ee}{\end{equation}}
\newcommand{\bea}{\begin{eqnarray}}
\newcommand{\eea}{\end{eqnarray}}

\newcommand{\Eq}[1]{Eq.\,(\ref{#1})}
\newcommand{\Eqs}[1]{Eqs.\,(\ref{#1})}

\newcommand{\la}{\langle}
\newcommand{\ra}{\rangle}
\newcommand{\dg}{\dagger}

\begin{document}
\begin{CJK*}{GBK}{Song}

\title{ Quantum efficiency of charge qubit measurements
using a single electron transistor }

\author{Yin Ye $^{1}$, Jing Ping $^{1}$, HuJun Jiao$^{2}$,
Shu-Shen Li $^{1}$, and
Xin-Qi Li$^{3,1}$ \footnote{E-mail: lixinqi@bnu.edu.cn}  }
\address{$^1$ State Key Laboratory for Superlattices and
Microstructures, Institute of Semiconductors, Chinese Academy of
Sciences, P.O.~Box 912, Beijing 100083, China}
\address{$^2$ Department of Physics, Shanxi University,
         Taiyuan 030006, China }
\address{$^3$ Department of Physics, Beijing Normal University,
Beijing 100875, China }


\date{\today}

\begin{abstract}
The quantum efficiency, which characterizes the quality of information
gain against information loss, is an important figure of merit for any
{\it realistic} quantum detectors in the {\it gradual}
process of collapsing the state being measured.
In this work we consider the problem of solid-state charge qubit
measurements with a single-electron-transistor (SET).
We analyze two models: one corresponds to a strong
response SET, and the other is a tunable one in response strength.
We find that the response strength would essentially bound the quantum
efficiency, making the detector non-quantum-limited.
Quantum limited measurements, however, can be achieved in the limits
of strong response and asymmetric tunneling.
The present study is also associated with appropriate justifications
for the measurement and backaction-dephasing rates, which were
usually evaluated in controversial methods.
\end{abstract}

\maketitle


In quantum mechanics the Copenghagen's postulate assumes that
a quantum measurement would instantaneously collapse the state
being measured onto one of the eigenstates of the observable.
This is the concept of perfect projective measurement.
In practice, however, any realistic quantum detectors cannot
realize such type of measurements.
Actually, a realistic quantum measurement is a process of
{\it gradually} collapsing the state being measured.
In this context, the {\it quantum efficiency}
is an important figure of merit for a quantum detector.
To be more specific, let us consider the measurement of
a two-state (qubit) system.
Assuming the qubit is in an idle state which is simply
a superposition of the logic basis states,
but experiences no rotational operation between them.
Further, we focus on a quantum measurement using the
so-called quantum non-demolition (QND) detector,
which only dephases the quantum coherence defined by the superposition,
but does not flip the basis states.
This situation coincides with the fact that the measurement operator
is commutable with the qubit Hamiltonian, which is one of the major
criterions of the QND measurement in general \cite{Bra92,Bra96}.
While for more general case the QND measurements of a qubit
were discussed in Refs.\ \cite{Ave02,But05},
we restrict us in the present work to a simpler case as mentioned above
with, however, a particular interest in the quantum efficiency.

Now, consider the qubit measurements with a QND detector.
During the (gradual) collapse process, one can get the measurement result
only {\it after} some time until the signal-to-noise ratio reaches unity,
owing to the stochastic nature of the elementary events leading to
the collapse (such as tunneling or excitations in the detector).
This consideration actually defines a {\it measurement} time ($\tau_m$).
On the other hand, the qubit state being measured would be
inevitably dephased because of the detector's backaction,
from which we can define a {\it dephasing} rate ($\Gamma_d$).
Then, the {\it quantum efficiency} of a realistic detector
is defined by $\eta=1/(2\Gamma_d\tau_m)$, which can also
be related with an interpretation of information gain
{\it versus} information loss \cite{CGS03,Ave05,Kor08}.
We say that a quantum detector performs quantum limited measurements
when $\eta$ reaches unity.

In this work we consider a concrete realization of qubit measurements,
say, a solid-state charge qubit measured
with a single electron transistor (SET) \cite{Ave86,Ful87,Mei90,Dev00}.
When the tunnel coupling between the charge states is quenched,
the measurement falls into the scenario of QND type.
Moreover, it is believed that the SET detector holds promising
applications in solid-state quantum computation \cite{Dev00,Sch01},
whose measurement properties have been therefore received considerable
attention in the past years \cite{Kor01,Moz04,Ave01,vdB02,Gur05,Oxt06,Li09}.
While in the higher-order cotunneling regime the measurement can reach the
quantum limit ($\eta=1$) in principle \cite{Ave01,vdB02}, it was found that
the quantum efficiency of the SET detector is rather poor in the weak response
and sequential tunneling regime \cite{Sch01,Kor01,Moz04}.

More recently, however, it was found that the signal-to-noise ratio
in the power spectrum of qubit oscillation measurements
using the SET, another important figure of merit,
can reach and even exceed the {\it ideal} value of quantum limited
linear-response detectors \cite{Li09},
i.e., the Korotkov-Averin bound \cite{K-A01}.
To achieve such a result, the necessary conditions for the SET detector
are an asymmetric tunnel coupling with the leads
and a strong-response to the qubit.
This result, together with some earlier investigations \cite{Gur05,Oxt06},
provides a hint that the SET detector seems able to perform
ideal quantum-limited measurement under similar conditions.
In this paper, rather than the continuous measurement of qubit oscillations
\cite{Gur05,Li09}, we will focus on the QND-type measurements for the idle
(superposition) state of a qubit, to show how the quantum efficiency
depends on the response strength and tunnel coupling to the leads
and how a quantum-limited measurement can be achieved.
The study will be associated with appropriate justifications
for the measurement and backaction-dephasing rates,
which were usually evaluated in controversial methods.

\begin{figure}[h]
\begin{center}
\includegraphics[width=7cm]{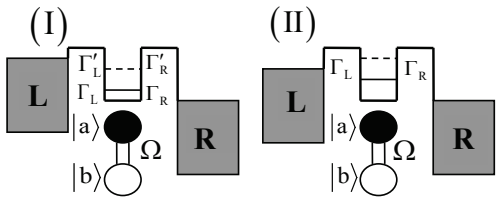}
\caption{\label{fig1} Schematic setup of using single electron
transistor to measure a solid-state qubit.
Model (I) stems from the metallic SET, while model (II)
can be realized with the semiconductor nanostructure
of quantum dot SET.  }
\end{center}
\end{figure}

\vspace{0.2cm}
{\it Model Description}.---
Consider a charge qubit, say, an electron in a pair of coupled
quantum dots with degenerate dot levels $E_a=E_b$
and inter-dot coupling amplitude $\Omega$,
measured by a single electron transistor.
For the qubit, its Hamiltonian reads $H_{\rm qu}=\sum_{j=a,b}E_j
|j\ra \la j| + \Omega (|a\ra \la b| +{\rm H.c.})$.
For the SET, in this work we consider two models\cite{Li09},
as schematically shown in Fig.\ 1.
For both models,
the qubit state can be discriminated from the different currents
flowing through the SET detector, which are correlated
with the qubit states $|a\ra$ and $|b\ra$
owing to the Coulomb interaction $H_{\rm int}=U n_c n_a$.
Here, $n_c$ and $n_a$ are the number operators of
the SET central dot and the qubit target state $|a\ra$.

In practice, there exist {\it metallic} and {\it semiconductor}
realizations for single electron transistors.
For semiconductor-quantum-dot-based SET, the models schematically
shown in Fig.\ 1 can be understood straightforwardly, by noting
that the dot levels are discrete and assuming only one level
in the voltage window. Accordingly, there are only
two charge states, say, an empty and single-electron-occupation states.
In the metallic case \cite{Sch01}, the central dot is an island
with dense energy levels.
However, owing to the strong Coulomb blockade effect, it can still
involve only two charge states in the transport process.
Formally, these two charge states together with their couplings to the electrodes
can be described similarly as the semiconductor quantum-dot SET.
Particular treatments based on Fermi's golden rule for the {\it metallic}
island-electrode coupling rates are referred to Ref.\ \cite{Sch01}.

In our present work, we do not specify what type of SET is concerned,
but instead present a unified study for two different models,
i.e., (I) and (II) in Fig.\ 1.
In model (I), the SET dot level is always within the voltage window,
regardless the qubit state in $|b\ra$ or $|a\ra$,
but with different coupling strengths to the leads,
i.e., $\Gamma_{L(R)}$ and $\Gamma'_{L(R)}$.
In model (II), the SET dot level is within the voltage window only
for qubit in state $|b\ra$. It locates outside the voltage window
if qubit in state $|a\ra$, due to the stronger Coulomb interaction
with the qubit.
While model (II) is purely a strong response detector,
we parameterize the response strength of model (I) as follows:
$\Gamma_L(\Gamma_L^\prime)=(1\pm\xi)\bar{\Gamma}_L$,
$\Gamma_R(\Gamma_R^\prime)=(1\pm\zeta)\bar{\Gamma}_R$,
and $\gamma=\bar{\Gamma}_R/\bar{\Gamma}_L$.
Then, $\xi$ and $\zeta$ characterize
the response strengths of the detector to the qubit,
while $\bar{\Gamma}_{L(R)}=(\Gamma_{L(R)}+\Gamma'_{L(R)})/2$
describe the average couplings.
These rates, originally, are related with a number of microscopic quantities,
such as the tunnel-coupling amplitudes and the density-of-states of the leads
through the Fermi's golden rule \cite{Sch01}.
In particular, the above identification of the rates is also associated
with the consideration of large-bias voltage and low temperatures,
which make the Fermi function be unity.

%
%

\vspace{0.2cm}
{\it Conditional Master Equation.}---
Following Ref.\ \cite{Li09} and referring to the Appendix A
of the present work for more details, a number-conditioned
master equation can be formally expressed as
\begin{align}\label{Eq2}
\dot{\rho}^{(n)}=&-i\mathcal{L}\rho^{(n)}
-\sum_{j=0,\pm 1} \mathcal{R}_j \rho^{(n+j)} .
\end{align}
Here, $\rho^{(n)}$ describes the state of the {\it system},
say, the qubit plus the SET dot,
conditioned on the electron number ``$n$"
tunnelled through the SET detector
(i.e., the number of electrons arrived to the right electrode).
The Liouvillian $\mathcal{L}$ is a commutator related to the
system Hamiltonian.
The superoperators $\mathcal{R}_j$ describe electron transfer
between the SET dot and leads, with explicit forms
given in Ref.\ \cite{Li09}.
To implement this number-conditioned master equation in practice,
we need to specify the state basis. For the both models described above,
we choose the same state basis: $|1\rangle\equiv|0a\rangle$,
$|2\rangle\equiv|0b\rangle$, $|3\rangle\equiv|1a\rangle$, and
$|4\rangle\equiv|1b\rangle$. In this notation $|0(1)a(b)\rangle$
means that the SET dot is empty (occupied) and the qubit is in state
$|a(b)\rangle$.
Explicitly, for model (I) we have \cite{Li09}
\begin{subequations}\label{nME-1}
\begin{align}
\dot{\rho}^{(n_R)}_{11}=&-\Gamma_L^\prime\rho^{(n_R)}_{11}
    +\Gamma_R^\prime\rho^{(n_R-1)}_{33}\\
\dot{\rho}^{(n_R)}_{33}=&\Gamma_L^\prime\rho^{(n_R)}_{11}
    -\Gamma_R^\prime\rho^{(n_R)}_{33}\\
\dot{\rho}^{(n_R)}_{22}=&-\Gamma_L\rho^{(n_R)}_{22}
    +\Gamma_R\rho^{(n_R-1)}_{44}\\
\dot{\rho}^{(n_R)}_{44}=&\Gamma_L\rho^{(n_R)}_{22}
    -\Gamma_R \rho^{(n_R)}_{44}\\
\dot{\rho}^{(n_R)}_{12}=&-\frac{\Gamma_L+\Gamma_L^\prime}{2}\rho^{(n_R)}_{12}
 +\frac{\Gamma_R+\Gamma_R^\prime}{2}\rho^{(n_R-1)}_{34}\\
\dot{\rho}^{(n_R)}_{34}=&-iU\rho^{(n_R)}_{34}+\frac{\Gamma_L
   +\Gamma_L^\prime}{2}\rho^{(n_R)}_{12}
   -\frac{\Gamma_R+\Gamma_R^\prime}{2}\rho^{(n_R)}_{34}  ,
\end{align}
\end{subequations}
while for model (II) it reads
\begin{subequations}\label{nME-2}
\begin{align}
\dot{\rho}^{(n_R)}_{11}=&\Gamma_L\rho^{(n_R)}_{33}+\Gamma_R\rho^{(n_R-1)}_{33}\\
\dot{\rho}^{(n_R)}_{33}=&-(\Gamma_R+\Gamma_L)\rho^{(n_R)}_{33}\\
\dot{\rho}^{(n_R)}_{22}=&-\Gamma_L\rho^{(n_R)}_{22}+\Gamma_R \rho^{(n_R-1)}_{44}\\
\dot{\rho}^{(n_R)}_{44}=&\Gamma_L\rho^{(n_R)}_{22}-\Gamma_R \rho^{(n_R)}_{44}\\
\dot{\rho}^{(n_R)}_{12}=&-\frac{\Gamma_L}{2}\rho^{(n_R)}_{12}+
\frac{\Gamma_L}{2}\rho^{(n_R)}_{34}+\Gamma_R\rho^{(n_R-1)}_{34}\\
\dot{\rho}^{(n_R)}_{34}=&-iU \rho^{(n_R)}_{34}+\frac{\Gamma_L}{2}
\rho^{(n_R)}_{12} - (\Gamma_R+\frac{\Gamma_L}{2}) \rho^{(n_R)}_{34}  .
\end{align}
\end{subequations}
In these equations, $\rho^{(n_R)}$ means that the state is conditioned
on the number of electrons tunneled through the {\it right} junction
of the SET, which is more explicitly labeled here by ``$n_R$",
instead of ``$n$" in \Eq{Eq2}.
Also, differing from \Eq{Eq2} which supports a bidirectional transport
($j=+1$ and $j=-1$), in \Eqs{nME-1} and (\ref{nME-2}) only the forward
unidirectional process ($j=-1$) survives.
This is owing to the fact that the measurement is performed
in a large-bias determined sequential tunneling regime.
In this case, the SET-dot level that supports the transport current
is deeply embedded between the Fermi levels of the two leads,
leading thus to a strong suppression of the backward process.
Finally, as explained in the part of introduction, we are
interested in a QND-type measurement which allows us to
set $\Omega=0$ in \Eqs{nME-1} and (\ref{nME-2}).

\vspace{0.2cm}
{\it Measurement and Dephasing Rates}.---
The measurement time $\tau_m$, in realistic gradual collapse
quantum measurement, is the average time needed to filter out
the measurement signal from detector's noisy output.
For model (I), it can be determined by a technique
of wave-packet analysis \cite{Sch01}, which is actually
equivalent to the counting-statistics technique employed
in Ref.\ \cite{Ave05} to analyze the measurement rate
with a strong response point-contact detector.
Discrete-Fourier-transforming \Eq{nME-1} in terms of
$\rho(k,t)=\Sigma^{\infty}_{n=0}\rho^{(n)}(t)e^{i nk}$,
with $k\in[0,2\pi]$, yields
\begin{subequations}\label{kME-1}
\begin{align}
\dot{\rho}_{11}=&-\Gamma_L^\prime\rho_{11}+\Gamma_R^\prime\rho_{33}e^{ik}\\
\dot{\rho}_{33}=&\Gamma_L^\prime\rho_{11}-\Gamma_R^\prime\rho_{33}\\
\dot{\rho}_{22}=&-\Gamma_L\rho_{22}+\Gamma_R\rho_{44}e^{ik}\\
\dot{\rho}_{44}=&\Gamma_L\rho_{22}-\Gamma_R \rho_{44}\\
\dot{\rho}_{12}=&-\frac{\Gamma_L+\Gamma_L^\prime}{2}\rho_{12}+
\frac{\Gamma_R+\Gamma_R^\prime}{2}\rho_{34}e^{ik}\\
\dot{\rho}_{34}=&-iU\rho_{34}+\frac{\Gamma_L+\Gamma_L^\prime}{2}\rho_{12}-
\frac{\Gamma_R+\Gamma_R^\prime}{2}\rho_{34}
\end{align}
\end{subequations}
These equations can be split into three groups, i.e.,
Eqs.\ (4a)-(4b), Eqs.\ (4c)-(4d), and Eqs.\ (4e)-(4f).
Considering the characteristic solution proportional to
$e^{i\omega t}$, for each group we can obtain two eigenvalues.
Since for model (I) we are able to distinguish (read out) the qubit
state only after relatively large number of electrons transmitted
through the SET so that a measurement current is well defined,
we need thus a solution only for small $k$, which dominantly
contributes to the electron-number distribution function.
Then, following Ref.\ \cite{Sch01},
in the limit of $k\ll 1$, from Eqs.\ (4a)-(4b) and Eqs.\ (4c)-(4d),
we can carry out the smallest two eigenvalues, formally expressed as
$\omega_{\mu}(k)=v_{\mu}k+\frac{1}{2}i v_{\mu}f^{\mu}k^{2}$,
with $\mu=a$ and $b$ corresponding to the qubit states.
Here, $v_{\mu}$ are the wave-packet group velocities, being
identical to the stationary currents $I_{\mu}$ associated
with the qubit state $|\mu\rangle$,
while $f^{\mu}$ are the respective Fano factors.
The measurement time then reads \cite{Sch01},
$\tau_{\mathrm{m}}=\left(\sqrt{f^{a}v^{a}}
                 +\sqrt{f^{b}v^{b}}\right)^2
   /  \left(v^{a}-v^{b}\right)^{2}$.

For model (II), which represents a strong-response measurement,
instead of the wave-packet analysis described above,
a better way to determine $\tau_m$ is expected by the
following consideration.
Since for qubit in state $|a\ra$ no electron
can tunnel through the SET and arrives at the right lead,
one then immediately knows the qubit in state $|b\ra$
as soon as an electron is detected at the right lead.
This is nothing but a single-shot measurement completed,
and thus the measurement time is the average
waiting time for such a tunneling event.
Applying the method of waiting-time analysis \cite{Wel08},
in Appendix A we present some details of calculating the
measurement time, and obtain $\tau_{\rm m}=1/\Gamma_L+1/\Gamma_R$.

Now we discuss how to {\it reliably} determine the measurement
backaction dephasing rate $\Gamma_d$, for the present strong
coupling models (I) and (II)
(i.e. the large-$U$ interaction between the SET and qubit).
The underlying complexity originates from the fact that
the off-diagonal element of $\rho(t)$,
i.e., the simplest measure of qubit coherence,
does {\it not} decay with single exponential rate in our case.
Then, we may determine $\Gamma_d$ differently as follows:

{\it (i)}
Regarding the slowest decay rate as the dephasing rate
leads to $\Gamma_{d}=\mathrm{min}[\mathrm{Im}\omega_{\mu}]$,
with $\omega_{\mu}$ the two eigenvalues of  Eqs.\ (4e)-(4f).
Since this definition ignores the weight of each
exponentially decaying component,
it breaks down as the weight of the
{\it slowest} decay component is the {\it smallest} one.

{\it (ii)}
Borrowing a technique in quantum optics, $\Gamma_d$ can be defined
better from the Glauber coherence function \cite{Oxt06}:
$g(\tau)=\langle [\sigma^{\dagger}(\tau)\sigma(0)
  + {\rm h.c.} ] \rangle_{ss}$,
where $\sigma^{\dagger}=|a\rangle\langle b|$,
$\sigma^{\dagger}(\tau)$ is the operator in Heisenberg picture,
and $\langle \cdots \rangle_{ss}$ means average
with respect to the steady-state.
Then, one defines
$\Gamma^{-1}_{d} =\int_{0}^{\infty}g(\tau)d\tau$.
As we will see later, this definition suffers also to some extent
the drawback of the method {\it (i)} above.

{\it (iii)}
In this work, in similar spirit of {\it (ii)},
we propose to determine $\Gamma_d$ simply as:
$\Gamma^{-1}_{d}= \int_{0}^{\infty}
 \rho_{ab}(\tau)d\tau  /  \rho_{ab}(0)$.
Differently, however, we assume and emphasize that at the beginning
of measurement the SET dot is empty.
We will see below that this identification
is important, particularly for small $\Gamma_R$,
compared to the above Glauber coherence function method.

\begin{figure}[h]
\begin{center}
\includegraphics[width=8cm]{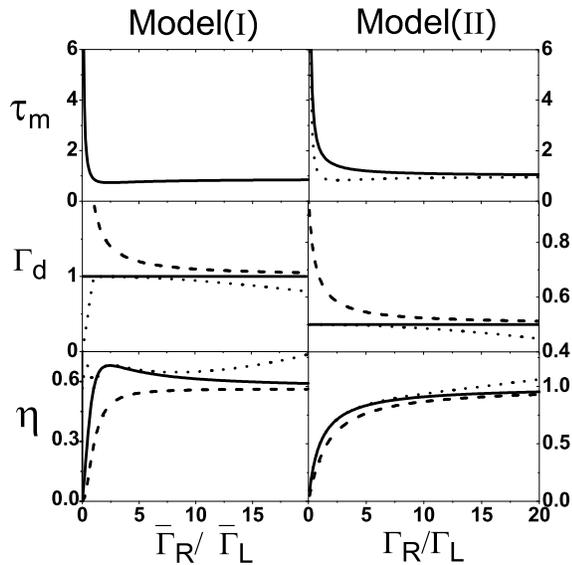}
\caption{\label{fig3}
Measurement time $\tau_m$, backaction dephasing rate $\Gamma_d$,
and quantum efficiency $\eta=1/(2\Gamma_d\tau_m)$,
for models (I) and (II).
For $\tau_m$ as shown in the upper panels,
in addition to results from the wave-packet
method for model (I) (solid line) and (II) (dotted line),
the more desired result of model (II) is based on the waiting-time
analysis (solid line).
Plotted in the mid-panels are results of $\Gamma_d$ from, respectively,
decay of $\rho_{ab}(\tau)$ with empty SET dot initially
(solid line), decay of the Glauber correlation function (dashed line),
and the smallest eigenvalue method (dotted line).
Accordingly, the respective efficiency is plotted in the lower
panels with the same type of lines.
Parameters: Taking the SET dot level associated with the qubit
state $|b\ra$ as the reference of zero energy,
and $\bar{\Gamma}_L$ and $\Gamma_L$ as the energy units,
we set
$\mu_L=50$, $\mu_R=-20$, and $U=40$ for model (I)
while $U=60$ for model (II). And, for model (I),
the response strength $\xi=\zeta=0.9$. }
\end{center}
\end{figure}

\vspace{0.2cm}
{\it Numerical Results and Discussions.}---
In Fig.\ 2 we plot $\tau_m$, $\Gamma_d$,
and the quantum efficiency $\eta=1/(2\Gamma_d\tau_m)$
against the setup asymmetry, respectively,
for models (I) and (II).
For the result of model (I),
shown by the left-column three panels of Fig.\ 2,
the measurement time $\tau_m$ is carried out using the
wave-packet analysis, while the dephasing rate $\Gamma_d$ is
computed respectively by the three methods {\it (i)}-{\it (iii)}
outlined above.
For $\Gamma_d$, we observe
serious deviation from the desired result (solid line),
indicating that either the smallest-eigenvalue method (dotted line)
or the Glauber coherence function approach (dashed line)
breaks down in this context,
particularly for small $\bar{\Gamma}_R/\bar{\Gamma}_L$.
Since we are considering a strong coupling detector
(i.e. with large $U$ interaction), the dephasing rate is dominantly
determined by $\bar{\Gamma}_L$, being nearly independent of
$\bar{\Gamma}_R/\bar{\Gamma}_L$.
Concerning the efficiency of model (I), we find that merely increasing
the tunneling asymmetry $\bar{\Gamma}_R/\bar{\Gamma}_L$ cannot reach
the quantum limit in weak or even relatively strong response regimes.
In addition to the tunneling asymmetry,
the efficiency is also bounded by, and actually quite sensitive to,
the response strength. This feature contradicts our intuition
that an asymmetric SET is like a point-contact (PC) detector, by noting that
an ideal PC detector has unit quantum efficiency which is independent of
the response strength.
We also notice that this conclusion differs from some statements
in Ref.\ \cite{Oxt06},
but agrees well with the fact that the signal-to-noise ratio
is considerably affected by the response strength \cite{Li09,Li07}.

For the strong response detector model (II),
we now demonstrate that the quantum limit can be achieved.
The respective $\tau_m$, $\Gamma_d$ and $\eta$ are plotted
in the right column of Fig\ 2.
The solid-line in the upper panel
plots the simple result $\tau_m=1/\Gamma_L+1/\Gamma_R$
from the waiting-time analysis,
while the dotted-line shows the result from
the wave-packet analysis as a comparison.
For $\Gamma_d$, shown in the mid-panel,
we again find that the dephasing is dominantly
caused by the tunneling events from the left lead of SET
to its central dot, i.e., $\Gamma_d=\Gamma_L/2$.
The dotted and dashed lines, respectively from the smallest eigenvalue
and the Glauber coherence function methods,
clearly show the extent of failure when applied to this model.
Therefore, quite straightforwardly, the quantum efficiency
is analytically obtained as:
\begin{align}
\eta=\Gamma_R/(\Gamma_L+\Gamma_R) .
\end{align}
From this result, we see that the measurement is to be
quantum limited with increasing $\Gamma_R/\Gamma_L\gg 1$.

Note that what this model (II) performs is actually a {\it strong projective
measurement} since one can immediately conclude that the qubit is
in one of the basis states if an electron tunnels through the detector.
In terms of ensemble average times as used in the quantum efficiency definition
(i.e., average over large number of transmitted and reflected events),
the random tunneling through the left junction (equivalently,
the random occupation of the SET dot) determines the dephasing rate
of the qubit, while the detection in the right reservoir
for the tunneled electron determines the measurement time.
In this ensemble average sense, the qubit coherence is
destroyed by the first jump, and the measurement rate is governed
by the current $\Gamma_L \Gamma_R/(\Gamma_L+\Gamma_R)$.

For strong projective measurement, however, it may not be immediately
clear how the fact whether the detector is quantum-limited or not
would manifest itself in each single realization of measurement.
We may explain this issue in terms of possible information
loss in each single realization of projective measurement.
Let us reconsider the measurement process. That is, the left reservoir
(source) emits a measuring electron, this electron first tunnels through
the left junction, then through the right one, and is finally detected in
the right reservoir (drain). From an information-theoretic point of view,
the right-hand-side detector cannot distinguish whether there is a tunneling
event from the left reservoir to the SET dot.
This implies an information loss, and results in a non-unit quantum efficiency
which makes sense even at the level of single projective measurement.
To avoid this information loss, one possible way is to detect the tunneling
electron in the right reservoir with increase of the tunneling
asymmetry $\Gamma_R/\Gamma_L$, as indicated by the above calculation.
Interestingly, based on the information-theoretic discussion,
for model (II) with arbitrary $\Gamma_R/\Gamma_L$,
one can also reach the quantum limit by alternatively detecting
electron tunneling through the left junction.
This can be implemented by adding a nearby point-contact
to detect the charge state of the SET dot,
similar to the experiment by Gustavsson {\it et al} \cite{Gus06}.
In this way, the information-gain rate coincides with
the dephasing rate, leading thus to a unit quantum efficiency.

\vspace{0.2cm}
{\it Conclusion.}---
To summarize, we have presented an analysis for the quantum efficiency
of qubit measurements with a SET detector.
Two models were analyzed: one corresponds to a strong
response SET, and the other is a tunable one in response strength.
We found that quantum limited measurements can be achieved only in a
limiting case of strong response as schematically shown by model (II),
together with an asymmetric tunneling setup.
In the most range of response strengths, the SET detector cannot reach
the quantum limit of efficiency.
These results can be qualitatively understood by means of an
information-theoretic interpretation,
and are quantitatively demonstrated with direct calculations
by appropriately justifying the measurement and backaction dephasing rates.

\vspace{0.3cm}
{\it Acknowledgements.}---
This work was supported by the NNSF of China under
grants No.\ 101202101 \& 10874176.


\appendix
\section{Waiting-Time Calculation}
%

Conditioned on the number of electrons tunneled through the SET,
rather than \Eq{Eq2}, we more explicitly present the number-conditioned
master equation as \cite{Li09}
\begin{align}\label{Eq2a}
 \dot{\rho}^{(n)}=&-i\mathcal{L}\rho^{(n)}-\frac{1}{2}\{[a_c^\dagger,
 A_{L}^{(-)}\rho^{(n)}-\rho^{(n)}A_{L}^{(+)}]\nl&
 +a_c^\dagger A_{R}^{(-)}\rho^{(n)}
 +\rho^{(n)}A_{R}^{(+)}a_c^\dagger
 \nl&-[a_c^\dagger\rho^{(n+1)}A_{R}^{(+)}+A_{R}^{(-)}
 \rho^{(n-1)}a_c^\dagger]+ {\rm H.c.}  \} .
\end{align}
Here, the Liouvillian $\mathcal{L}$ is defined by $\mathcal{L}(\cdots)=[H_S,\cdots]$,
where $H_S$ is the system (i.e. qubit plus SET-dot) Hamiltonian.
We introduce operators
$A_{\lambda}^{(\pm)}\equiv C_{\lambda}^{(\pm)}(\pm\mathcal{L})a_c$,
with $\lambda=L,R$ and $a_c^\dag$ $(a_c$)
the creation (annihilation) operator of the SET central dot.
The superoperoters $C_{\lambda}^{(\pm)}(\pm\mathcal{L})$
are the generalized spectral functions:
$C_{\lambda}^{(\pm)}(\pm\mathcal{L})=\int_{-\infty}^{+\infty}dt
C_{\lambda}^{(\pm)}(t)e^{\pm i\mathcal{L}t}$,
where the bath correlation functions
$C_{\lambda}^{(+)}(t)=\langle f_{c\lambda}^{\dg}(t)f_{c\lambda}\rangle_B$
and $C_{\lambda}^{(-)}(t)=\langle
f_{c\lambda}(t)f_{c\lambda}^{\dg}\rangle_B$.
In these correlators,
the average $\la \cdots \ra_B\equiv {\rm Tr}_B[(\cdots)\rho_B]$
is over the local thermal equilibrium state ($\rho_B$) of the SET leads,
and $f_{c\lambda}^{\dg}$ ($f_{c\lambda}$) are the leads creation (annihilation)
operators from the tunneling Hamiltonian
that describes the coupling to $a_c$ ($a_c^\dag$).

Notice that the usual unconditional state $\rho(t)$ is related to
$\rho^{(n)}(t)$ simply as: $\rho(t)=\sum_{n}\rho^{(n)}(t)$.
Based on \Eq{Eq2a}, the master equation about $\rho(t)$
can be readily obtained and rewritten as
\begin{align}
\dot{\rho}=-i\mathcal{L}\rho- \frac{1}{2} \left\{ \left[ \Pi_{L}
  -\Sigma^{(+)}_{L}-\Sigma^{(-)}_{L}\right]
  +\left[ \Pi_{R}-\Sigma^{(-)}_{R}\right]\right\} \rho .
\end{align}
In large bias limit (i.e. sequential tunneling limit),
the superoperators are defined through
\begin{align}
&\Pi_{L}\rho=\left[ a_{c}^{\dagger}A_{L}^{(-)}\rho
     +\rho A_{L}^{(+)}a_{c}^{\dagger}\right] +{\rm H.c.}    \nonumber\\
&\Pi_{R}\rho= a_{c}^{\dagger}A_{R}^{(-)}\rho+ {\rm H.c.}  \nonumber\\
&\Sigma^{(+)}_{L}\rho=a_{c}^{\dagger}\rho A_{L}^{(+)}+{\rm H.c.} \nonumber\\
&\Sigma^{(-)}_{L}\rho=A_{L}^{(-)}\rho a_{c}^{\dagger}+{\rm H.c.} \nonumber\\
&\Sigma^{(-)}_{R}\rho=A_{R}^{(-)}\rho a_{c}^{\dagger}+{\rm H.c.} \nonumber\\
\end{align}
In these equations, $A^{(\pm)}_{L(R)}=\Gamma_{L(R)} n^{(\pm)}_{L(R)}a_c$,
where $n^{(+)}_{L(R)}$ is the Fermi function of the left (right) lead,
while $n^{(-)}_{L(R)}=1-n^{(+)}_{L(R)}$.

Now we consider two counting schemes, and determine the respective
waiting times until the first tunneling event happens.
First, for a left-junction counting,
conditioned on the result that no electron
tunnels through it, the master equation reads
\begin{align}
\dot{\rho}=&-i\mathcal{L}\rho-\frac{1}{2}\left\{
   \left[ a_{c}^{\dagger}A_{L}^{(-)}\rho
   +\rho A_{L}^{(+)}a_{c}^{\dagger} + a_{c}^{\dagger}A_{R}^{(-)}\rho\right]
   +{\rm H.c.}\right \}
\end{align}
More explicitly, in the state basis $|1\rangle\equiv|0a\rangle$,
$|2\rangle\equiv|0b\rangle$, $|3\rangle\equiv|1a\rangle$, and
$|4\rangle\equiv|1b\rangle$ we have
\begin{subequations}
\begin{align}
\dot{\rho}_{11}=&0\\
\dot{\rho}_{22}=&-\Gamma_{L}\rho_{22}\\
\dot{\rho}_{33}=&-\Gamma_{L}\rho_{33}-\Gamma_{R}\rho_{33}\\
\dot{\rho}_{44}=&-\Gamma_{R}\rho_{44}
\end{align}
\end{subequations}
For an initially empty SET dot, i.e., $\rho(0)=|2\rangle \langle 2|$,
the solution reads $\rho_{11}(t)=\rho_{33}(t)=\rho_{44}(t)=0$,
and $\rho_{22}(t)=e^{-\Gamma_{L}t}$. Then, based on a quantum-jump concept
we know the probability that
the first electron tunnels through the left junction at time ``$t$",
$p(t)\propto {\rm Tr}\left[ \Sigma^{(+)}_{L} \rho(t) \right]
 =\Gamma_{L}e^{-\Gamma_{L}t} $.
Accordingly, the waiting time for the first tunneling event is
\begin{align}
\tau_m=\frac{\int_{0}^{\infty}dt ~t~ p(t)}{\int_{0}^{\infty}dt p(t)}
=\frac{1}{\Gamma_{L}} .
\end{align}

Second, for a right-junction counting and conditioned on the result
that no electron tunnels through it, but regardless of the situation
occurred at the left junction, the master equation reads
\begin{align}
\dot{\rho}=&-i\mathcal{L}\rho-\frac{1}{2}\left\{\left[
   a_{c}^{\dagger}A_{L}^{(-)}\rho +\rho A_{L}^{(+)}a_{c}^{\dagger}
   +a_{c}^{\dagger}A_{R}^{(-)}\rho \right.\right. \nl
   &  \left.\left.
   -a_{c}^{\dagger}\rho A_{L}^{(+)}
   -A_{L}^{(-)}\rho a_{c}^{\dagger}\right] +{\rm H.c.}  \right\}.
\end{align}
In the same state basis as above, we further have
\begin{subequations}
\begin{align}
\dot{\rho}_{11}=&\Gamma_{L}\rho_{33}\\
\dot{\rho}_{22}=&-\Gamma_{L}\rho_{22}\\
\dot{\rho}_{33}=&-\Gamma_{L}\rho_{33}-\Gamma_{R}\rho_{33}+\Gamma_{L}\rho_{11}\\
\dot{\rho}_{44}=&-\Gamma_{R}\rho_{44}+\Gamma_{L}\rho_{22} .
\end{align}
\end{subequations}
Also, for initial condition $\rho(0)=|2\rangle \langle 2|$,
the solution reads $\rho_{11}=0,
\rho_{22}=e^{-\Gamma_{L}t},\ \rho_{33}=0,
\ {\rm and}~
\rho_{44}=(e^{-\Gamma_{R}t}-e^{-\Gamma_{L}t})\Gamma_L/(\Gamma_L-\Gamma_R)$.
Then,
$p(t)\propto {\rm Tr}\left[ \Sigma^{(-)}_{R} \rho(t) \right]
=(e^{-\Gamma_{R}t}-e^{-\Gamma_{L}t})\Gamma_L\Gamma_{R}/(\Gamma_L-\Gamma_R)$,
and the waiting time is obtained:
\begin{align}
\tau_m=\frac{\int_{0}^{\infty}dt~t~p(t)}{\int_{0}^{\infty}dtp(t)}
=\frac{1}{\Gamma_{L}}+\frac{1}{\Gamma_{R}} .
\end{align}


\end{CJK*}
\end{document}